\documentstyle[12pt,epsf]{article}        
\renewcommand{\baselinestretch}{1.3}

 1

\catcode`\@=11 
\makeatletter
\def\@seccntformat#1{\csname the#1\endcsname.\hskip 1em}

\makeatother

\catcode`@=11
\def\chkspace{%
  \relax   
  \begingroup\ifhmode\aftergroup\dochksp@ce\fi\endgroup}
\def\dochksp@ce{%
  \unskip              
  \futurelet\chkspct@k\d@chkspc  
}
\def\d@chkspc{%
  \let\nxtsp@ce=\relax
  \ifx\chkspct@k.\else     
    \ifx\chkspct@k,\else
      \ifx\chkspct@k;\else
        \ifx\chkspct@k!\else
          \ifx\chkspct@k?\else
            \ifx\chkspct@k:\else
              \ifx\chkspct@k)\else
              \ifx\chkspct@k(\else
                \ifx\chkspct@k]\else
                  \ifx\chkspct@k-\else
                    \ifx\chkspct@k\egroup\else  
                      \let\nxtsp@ce=\put@space  
                    \fi
                  \fi
                \fi
              \fi
              \fi
            \fi
          \fi
        \fi
      \fi
    \fi
  \fi
  \nxtsp@ce
}
\def\put@space{$\;$}
\catcode`@=12

\def\ra{{$\rightarrow$}\chkspace}
\def\etal{{\it et al.}\chkspace}

\def\eg{{\it eg.}\chkspace}

\def\ep{{e$^+$e$^-$}\chkspace}

\def\gluino{\relax\ifmmode \tilde{g} \else $\tilde{g}$ \fi\chkspace}

\def\qq{\relax\ifmmode q\overline{q}
\else $q\overline{q}$ \fi\chkspace}

\def\bb{\relax\ifmmode b\bar{b}
       \else $b\bar{b}$ \fi\chkspace}
\def\ccrm{\relax\ifmmode {\rm c}\bar{\rm c}
       \else ${\rm c}\bar{\rm c}$ \fi\chkspace}
\def\cc{$c\bar{c}$ \chkspace}
\def\tt{\relax\ifmmode {\rm t}\bar{\rm t}
       \else ${\rm t}\bar{\rm t}$ \fi\chkspace}
\def\ss{\relax\ifmmode {\rm s}\bar{\rm s}
       \else ${\rm s}\bar{\rm s}$ \fi\chkspace}
\def\uu{\relax\ifmmode {\rm u}\bar{\rm u}
       \else ${\rm u}\bar{\rm u}$ \fi\chkspace}
\def\dd{\relax\ifmmode {\rm d}\bar{\rm d}
       \else ${\rm d}\bar{\rm d}$ \fi\chkspace}

\def\qqg{\relax\ifmmode q\overline{q}g
\else $q\overline{q}g$ \fi\chkspace}
\def\bbg{\relax\ifmmode b\overline{b}g
\else $b\overline{b}g$ \fi\chkspace}

\def\afb{\relax\ifmmode A_{FB} \else
{{$A_{FB}$}}\fi\chkspace}
\def\afbb{\relax\ifmmode A_{FB}^b \else
{{$A_{FB}^b$}}\fi\chkspace}
\def\pafb{\relax\ifmmode \tilde{A}_{FB} \else
{{$\tilde{A}_{FB}$}}\fi\chkspace}
\def\pafbb{\relax\ifmmode \tilde{A}_{FB}^b \else
{{$\tilde{A}_{FB}^b$}}\fi\chkspace}

\def\pafbzo{\relax\ifmmode \tilde{A}_{FB}|_{O(0)} \else
{{$\tilde{A}_{FB}|_{O(0)}$}}\fi\chkspace}
\def\pafbfo{\relax\ifmmode \tilde{A}_{FB}|_{\oalp} \else
{{$\tilde{A}_{FB}|_{\oalp}$}}\fi\chkspace}
\def\pafbso{\relax\ifmmode \tilde{A}_{FB}|_{\oalpsq} \else
{{$\tilde{A}_{FB}|_{\oalpsq}$}}\fi\chkspace}
\def\pafbto{\relax\ifmmode \tilde{A}_{FB}|_{\oalpc} \else
{{$\tilde{A}_{FB}|_{\oalpc}$}}\fi\chkspace}

\def\pafbbzo{\relax\ifmmode \tilde{A}_{FB}^b|_{O(0)} \else
{{$\tilde{A}_{FB}^b|_{O(0)}$}}\fi\chkspace}
\def\pafbbfo{\relax\ifmmode \tilde{A}_{FB}^b|_{\oalp} \else
{{$\tilde{A}_{FB}^b|_{\oalp}$}}\fi\chkspace}
\def\pafbbso{\relax\ifmmode \tilde{A}_{FB}^b|_{\oalpsq} \else
{{$\tilde{A}_{FB}^b|_{\oalpsq}$}}\fi\chkspace}
\def\pafbbto{\relax\ifmmode \tilde{A}_{FB}^b|_{\oalpc} \else
{{$\tilde{A}_{FB}^b|_{\oalpc}$}}\fi\chkspace}

\def\afbo0{\tilde{A}_{FB}|_{O(0)}}
\def\afbo1{\tilde{A}_{FB}|_{\oalp}}
\def\afbo2{\tilde{A}_{FB}|_{\oalpsq}}
\def\afbo3{\tilde{A}_{FB}|_{\oalpc}}

\def\lam{\relax\ifmmode \Lambda_{\overline{MS}}
       \else {{$\Lambda_{\overline{MS}}$}}\fi\chkspace}
\def\lamuds{\relax\ifmmode \Lambda^{(3)}_{\overline{MS}}
       \else {{$\Lambda^{(3)}_{\overline{MS}}$}}\fi\chkspace}
\def\lamudsc{\relax\ifmmode \Lambda^{(4)}_{\overline{MS}}
       \else $\Lambda^{(4)}_{\overline{MS}}$\fi\chkspace}
\def\lamudscb{\relax\ifmmode \Lambda^{(5)}_{\overline{MS}}
       \else $\Lambda^{(5)}_{\overline{MS}}$\fi\chkspace}

\def\alp{\relax\ifmmode \alpha_s\else $\alpha_s$\fi\chkspace}
\def\alpbar{\relax\ifmmode \bar{\alpha_s}
       \else $\bar{\alpha_s}$\fi\chkspace}
\def\alpmz{\relax\ifmmode \alpha_s(M_Z)\else $\alpha_s(M_Z)$\fi\chkspace}
\def\alpmzsq{\relax\ifmmode \alpha_s(M_Z^2)
       \else $\alpha_s(M_Z^2)$\fi\chkspace}

\def\oalp{\relax\ifmmode O(\alpha_s)\else{{O($\alpha_s$)}}\fi\chkspace}
\def\oalpsq{\relax\ifmmode O(\alpha_s^2)
           \else{{O($\alpha_s^2$)}}\fi\chkspace}
\def\oalpc{\relax\ifmmode O(\alpha_s^3)
           \else{{O($\alpha_s^3$)}}\fi\chkspace}
\def\oalpf{\relax\ifmmode O(\alpha_s^4)
           \else{{O($\alpha_s^4$)}}\fi\chkspace}

\def\rb{\relax\ifmmode R_3^b/R_3^{all}
           \else{{$R_3^b/R_3^{all}$}}\fi\chkspace}
\def\rc{\relax\ifmmode R_3^c/R_3^{all}
           \else{{$R_3^c/R_3^{all}$}}\fi\chkspace}
\def\ruds{\relax\ifmmode R_3^{uds}/R_3^{all}
           \else{{$R_3^{uds}/R_3^{all}$}}\fi\chkspace}
\def\ri{\relax\ifmmode R_3^i/R_3^{all}
           \else{{$R_3^i/R_3^{all}$}}\fi\chkspace}
\def\rj{\relax\ifmmode R_3^j/R_3^{all}
           \else{{$R_3^j/R_3^{all}$}}\fi\chkspace}
\def\alpi{\relax\ifmmode \alpha^i_s/\alpha^{all}_s
           \else{{$\alpha^i_s/\alpha^{all}_s$}}\fi\chkspace}

\def\mbz{\relax\ifmmode m_b(M_Z)
           \else{{$m_b(M_Z)$}}\fi\chkspace}
\def\mbb{\relax\ifmmode m_b(M_b)
           \else{{$m_b(M_b)$}}\fi\chkspace}

\def\plb{Phys. Lett.\chkspace}

\def\prd{Phys. Rev.\chkspace}

\def\z0{{$Z^0$}\chkspace}
\def\Dst{\relax\ifmmode {\rm D}^* \else {D$^*$}\fi\chkspace}
\def\Dpl{\relax\ifmmode {\rm D}^+ \else {D$^+$}\fi\chkspace}
\def\D0{\relax\ifmmode {\rm D}^0 \else {D$^0$}\fi\chkspace}
\def\Kst{\relax\ifmmode {\rm K}^* \else {K$^*$}\fi\chkspace}
\def\K0{\relax\ifmmode {\rm K}^0_s \else {K$^0_s$}\fi\chkspace}
\def\Kpl{\relax\ifmmode {\rm K}^+ \else {K$^+$}\fi\chkspace}
\def\Kstz{\relax\ifmmode {\rm K}^{*0} \else {K$^{*0}$}\fi\chkspace}

\def\beq{\begin{equation}}
\def\eeq{\end{equation}}
\def\bea{\begin{eqnarray}}
\def\eea{\end{eqnarray}}

\begin{document}

\thispagestyle{empty}
\begin{flushright}
{\renewcommand{\baselinestretch}{.75}
  SLAC--PUB--9206\\
  Revised July 2002\\
}
\end{flushright}

\vskip 1truecm
 
\begin{center}
{\large\bf
 AN IMPROVED STUDY OF THE STRUCTURE OF\\[-0.1cm]
 e$^+$e$^-$ \ra \bbg  EVENTS AND LIMITS ON THE ANOMALOUS \\[0.1cm]
 CHROMOMAGNETIC COUPLING OF THE $b$-QUARK$^*$
}

\end{center}
 
 
\begin{center}
 {\bf The SLD Collaboration$^{**}$}\\
Stanford Linear Accelerator Center \\
Stanford University, Stanford, CA~94309
\end{center}
 
\vspace{1cm}
 
\begin{center}
{\bf ABSTRACT }
\end{center}
 
\noindent
The structure of three-jet \ep \ra $\bbg$ events has been studied using 
hadronic
$Z^0$ decays recorded in the SLD experiment at SLAC.
Three-jet final states were selected and the CCD-based vertex detector was
used to identify two of the jets as $b$ or $\overline{b}$;
the remaining jet in each event was tagged as the gluon jet.
Distributions of the gluon energy and polar angle with respect to the electron
beam were measured over the full kinematic range, and used to test the
predictions of perturbative QCD.
We find that beyond-leading-order QCD calculations are needed to reproduce
the features seen in the data.
The energy distribution is sensitive to an anomalous $b$
chromomagnetic moment $\kappa$ at the $b\bar{b}g$ vertex.
We measured $\kappa$ to be consistent with zero and set 95\% C.L. limits on 
its value, 
$-0.06 < \kappa < 0.04$.

\vskip 1truecm

\vfill

\begin{center}
{\it 
Submitted to Physical Review D}
\end{center}

\vfill

{\footnotesize
$^*$ Work supported in part by Department of Energy
contract DE-AC03-76SF00515.}

\eject

\section{Introduction}

\noindent
The observation of $e^+e^-$ annihilation into final states containing three 
hadronic jets, and their interpretation in terms of the  process
$e^+e^-\rightarrow\qqg$~\cite{threejets}, provided the first direct  evidence
for the existence of the gluon, the gauge boson of the theory of  strong
interactions, Quantum Chromodynamics (QCD). 
In subsequent studies the jets  were usually energy ordered, and the
lowest-energy jet was assigned as the gluon;  this is correct roughly 80\% of
the time, but preferentially selects low-energy gluons.
If the gluon jet can be tagged explicitly, event-by-event, the full kinematic
range of gluon energies can be explored, and more detailed tests of QCD can
be performed~\cite{BO}.
Due to advances in  vertex-detection this is now possible using \ep \ra $\bbg$
events.
The large mass  and relatively long lifetime, $\sim$ 1.5 ps, of the leading $B$
hadron in $b$-quark  jets~\cite{quark} lead to decay signatures that 
distinguish
them from lighter-quark ($u$, $d$, $s$ or $c$) and gluon jets.
We used our charge-coupled-device (CCD)-based vertex detector 
(VXD)~\cite{VXD3} to identify in each
event the two jets that contain the $B$  hadrons, and hence to tag the gluon
jet.
This allowed us to measure the gluon energy and polar-angle distributions over
the full kinematic range.

Additional motivation to study the \bbg system has been provided by 
measurements involving inclusive \z0\ra\bb decays.
Several early determinations~\cite{electrow} of $R_b$ =
$\Gamma(Z^0\rightarrow$\bb)/$\Gamma(Z^0\rightarrow$\qq)  
differed from Standard Model (SM)
expectations at the few standard deviation level. 
More recently it has been noted that the LEP measurement of the
$b$-quark forward-backward asymmetry, $A_{FB}^b$, lies roughly 
2.5 standard deviations below the SM expectation.
Since one expects new high-mass-scale  dynamics to couple to the massive
third-generation fermions, these  measurements have aroused considerable 
interest and speculation. 
We have therefore investigated in detail the strong-interaction dynamics of the
$b$-quark.
We have compared the strong coupling of the gluon to $b$-quarks with that to
light- and charm-quarks~\cite{sldflav}, as well as tested parity (P) and
charge$\oplus$parity (CP) conservation at the \bbg  vertex~\cite{sldsymm}.
We have also studied    the structure of \bbg events, via the  distributions of
the gluon  energy and polar angle with respect to (w.r.t.) the
beamline~\cite{sldbbg}, using the JADE algorithm~\cite{jade}
for jet definition.

Here we update the \bbg structure measurements using a data
sample more than 3 times larger than in our earlier study, and recorded 
in the upgraded vertex detector, which allowed us to improve
significantly the gluon-jet tagging.
In addition we extended our study to include the Durham, Geneva, E, E0 
and P algorithms~\cite{sldalp} to define jets,
and compared these results with perturbative QCD predictions.
This constitutes a more detailed test of QCD and enabled us to study
systematic effects arising from the jet definition algorithm.

Furthermore, we have used these data to study possible deviations from 
QCD in the form of radiative corrections induced by new physics,
in terms of the $b$-quark chromomagnetic moment.
In QCD this is induced at the one-loop
level and is of order $\alpha_s$/$\pi$. 
A more general $\bbg$ Lagrangian term with a modified coupling~\cite{tom1}
may be written:  
\begin{equation}
{\cal L}^{b\overline{b}g} =  g_s\overline{b}T_a \{ \gamma_{\mu} + 
\frac{i\sigma_{\mu\nu}k^{\nu}}{2m_b}(\kappa - i \tilde{\kappa}\gamma_5)\} 
bG_a^{\mu},\label{lag}
\end{equation}
\noindent
where $\kappa$ and $\tilde{\kappa}$ parametrize the anomalous chromomagnetic
and chromoelectric moments, respectively, which might arise from physics beyond
the SM. 
The effects of the chromoelectric moment are sub-leading w.r.t. those of the
chromomagentic moment, so for convenience we set $\tilde{\kappa}$ to zero.
A non-zero $\kappa$ would be observable as a modification~\cite{tom1} of the
gluon energy distribution in $\bbg$ events relative to the standard QCD case.
We have used our precise measurements of the gluon energy distributions to
set the most stringent limits on $\kappa$.

\section{$b\bar{b}g$ Event Selection}

We used hadronic decays of $Z^0$ bosons produced by $e^+e^-$ annihilations at
the SLAC Linear Collider (SLC) and recorded in the SLC Large Detector
(SLD)~\cite{SLD}. 
The criteria for selecting hadronic \z0 decays and the charged tracks used  for
flavor-tagging are described in~\cite{sldflav,dervan}.
Three-jet events were selected using iterative clustering algorithms applied
to the set of charged tracks in each event.
We used in turn the JADE, Durham, E, E0, P and Geneva algorithms.
The respective scaled-invariant-mass,
$y_{cut}$, values were chosen to maximise the statistical power of the
measurement, while keeping systematic errors small; they are
shown in Table~\ref{threejet}.

\begin{table}
\begin{center}
\begin{tabular}{|c|c|c|c|} \hline 
Jet algorithm  & $y_{cut}$ value & \# 3-jet events & efficiency\\
\hline
JADE   & 0.025   & 57341 & 12.2\% \\
Durham & 0.0095  & 46432 & 12.1\% \\
E      & 0.0275  & 66848 & 11.7\% \\
E0     & 0.0275  & 54163 & 11.2\% \\
P      & 0.02    & 60387 & 12.0\% \\
Geneva & 0.05    & 40895 & 12.8\% \\
\hline 
\end{tabular}
\end{center}
\caption{Number of selected 3-jet events, and gluon-jet tagging efficiency
(see text), for each algorithm. The statistical error on the efficiency
is roughly 0.04\%.}
\label{threejet}
\end{table}

Events classified as 3-jet states were retained if all three jets were well
contained within the barrel tracking system, with polar angle
$|\cos\theta_{jet}|$ $\leq$ 0.80. 
In addition, in order to select planar 3-jet events,
the sum of the angles between the jet axes was required to
be between 358 and 360 degrees.
From our 1996-98 data samples, comprising
roughly 400,000 hadronic \z0 decays, the numbers of selected events are shown
in Table~\ref{threejet}. 
In order to improve the energy resolution  the jet energies were rescaled
kinematically according to the angles between the jet axes, assuming energy and
momentum conservation and massless kinematics. 
The jets were then labelled in order of energy such that $E_1 > E_2 > E_3$.

Charged tracks with high quality information in the VXD as defined 
in~\cite{sldflav} were used to tag \bbg events.
For each such track we examined the impact-parameter, $d$, w.r.t. 
the interaction point (IP).
The resolution on $d$ is given by
$\sigma_d=$7.7$\oplus$29$/p\sin^{3/2}\theta$ $\mu$m in the plane transverse to
the beamline, and 9.6$\oplus$29$/p\sin^{3/2}\theta$ $\mu$m in any plane
containing the beamline, where $p$ is the track momentum in GeV/c, and
$\theta$ the polar angle, w.r.t. the beamline.

\begin{figure}[hbtp]
\begin{center}
\leavevmode
\epsfysize=15cm
\epsfbox{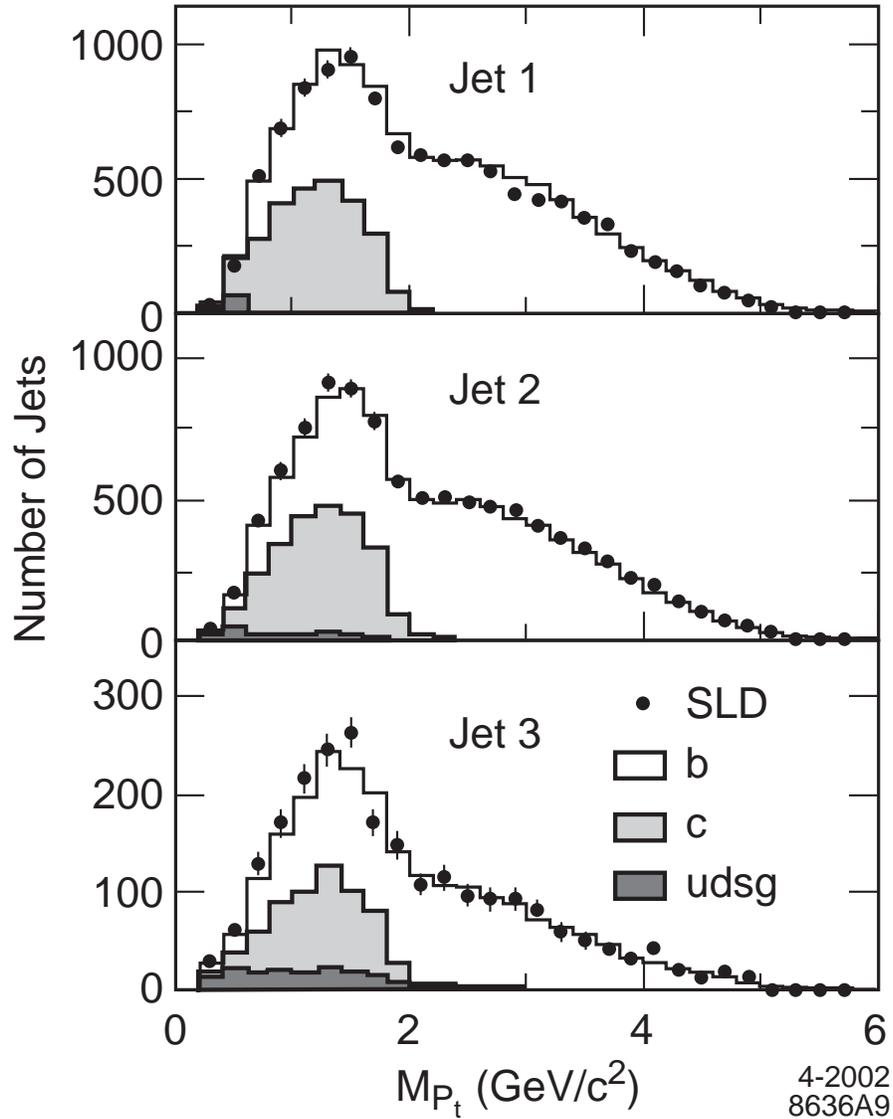}
\end{center}
\caption{The $M_{pt}$ distributions for vertices found in selected 3-jet
 events, defined using the JADE algorithm, 
labelled according to jet energy (dots); errors are statistical.
Histograms: simulated distributions for different jet flavors. Events were
selected by requiring that at least two jets contain a vertex, at least one
of which must satisfy $M_{pt}$ $>$ 2 GeV/$c^2$ (see text).
}
\label{mpt}
\end{figure}

Jets containing heavy hadrons were tagged using a topological
algorithm~\cite{davej} applied to the set of tracks associated with each jet.
A track density function was calculated, and a region of high total track 
density
well separated from the IP was identified as a 
vertex from the decay of a heavy hadron.
For each vertex, the $p_t$-corrected invariant mass~\cite{davej}, $M_{p_t}$,
 was calculated
using the set of tracks associated with the vertex, assuming the charged pion 
mass,
and the vertex axis, defined to be the vector from the IP to the 
reconstructed vertex position.
Fig.~\ref{mpt} shows the $M_{p_t}$ distributions
separately for vertices found in jets 1, 2 and 3 using, for illustration,
the JADE algorithm; results using the other algorithms (not shown)
are qualitatively similar.
The simulated contributions from true $b$, $c$, light and gluon jets are
indicated; the $c$, light and gluon jets populate predominantly the
region $M_{p_t}$ $<$ 2 GeV/$c^2$.
Events were retained in which exactly two jets contained such a vertex,
and at least one of them had $M_{p_t}$ $>$ 2 GeV/c$^2$.
In order to suppress events in which a single $B$-hadron decay gave
rise to two reconstructed vertices, 
the cosine of the angle between the
two vertex axes was required to be less than 0.7, and the distance
between the vertices, projected in a plane perpendicular to the beamline,
was required to exceed 0.12 cm. Roughly 1.1\% of the event sample was
rejected by these cuts.
In each selected event the jet without a vertex was tagged as the gluon jet.

For each algorithm, the number of tagged jets is shown in 
Table~\ref{jetpur};
also shown, in Table~\ref{threejet}, is the efficiency for tagging the 
gluon jet correctly in true \bbg events, which was calculated
using a simulated event sample generated with JETSET 7.4~\cite{jetset},
with parameter values tuned to hadronic $e^+e^-$ annihilation data
\cite{tuning}, combined with a simulation of $B$-decays tuned to
$\Upsilon$(4S) data~\cite{bdecay} and a simulation of the detector.
For the JADE algorithm, for example, 
the efficiency peaks at about 15\% for 18 GeV gluons.
Below 18 GeV the efficiency falls, to as low as 3\%, 
since lower-energy gluon jets are 
sometimes merged with the parent $b$-jet by the jet-finder.
Above 18 GeV the efficiency falls, to as low as 5\%, since
at higher gluon energies the correspondingly lower-energy $b$-jets are 
more difficult
to tag, and there is also a higher probability of losing a jet outside the
detector acceptance. Results for the other algorithms are qualitatively
similar.
The systematic error associated with the tagging efficiency was small and
was explicitly taken into account by the procedure for estimating 
systematic uncertainties that is described in Section~3. 

\begin{table}
\begin{center}
\begin{tabular}{|c|c|c|c|c|c|c|} \hline 
     & \multicolumn{2}{c|}{\hfil JADE \hfil}
     & \multicolumn{2}{c|}{\hfil Durham \hfil}
     & \multicolumn{2}{c|}{\hfil E \hfil} \\ \hline
jet label & \# jets & purity (\%) & \# jets & purity (\%) & \# jets & 
purity (\%) \\ \hline
3 & 4349 & 98.0 & 2952 & 97.0 & 5246 & 97.5 \\
2 & 740  & 90.2 & 890  & 92.4 & 1007 & 85.4 \\
1 & 150  & 71.0 & 138  & 73.4 & 148  & 70.8 \\
\hline\hline 
     & \multicolumn{2}{c|}{\hfil E0 \hfil}
     & \multicolumn{2}{c|}{\hfil P \hfil}
     & \multicolumn{2}{c|}{\hfil Geneva \hfil} \\ \hline
jet label & \# jets & purity (\%) & \# jets & purity (\%) & \# jets & 
purity (\%) \\ \hline
3 & 4027 & 98.0 & 4654 & 98.0 & 3491 & 93.9 \\ 
2 & 692  & 90.2 & 795  & 90.7 & 692  & 86.7 \\ 
1 & 151  & 70.7 & 155  & 72.1 & 181  & 63.4 \\ 
\hline
\end{tabular}
\end{center}
\caption{Tagging purities (see text).}
\label{jetpur}
\end{table}

\begin{figure}[hbtp]
\begin{center}
\leavevmode
\epsfysize=15 cm.
\epsfbox{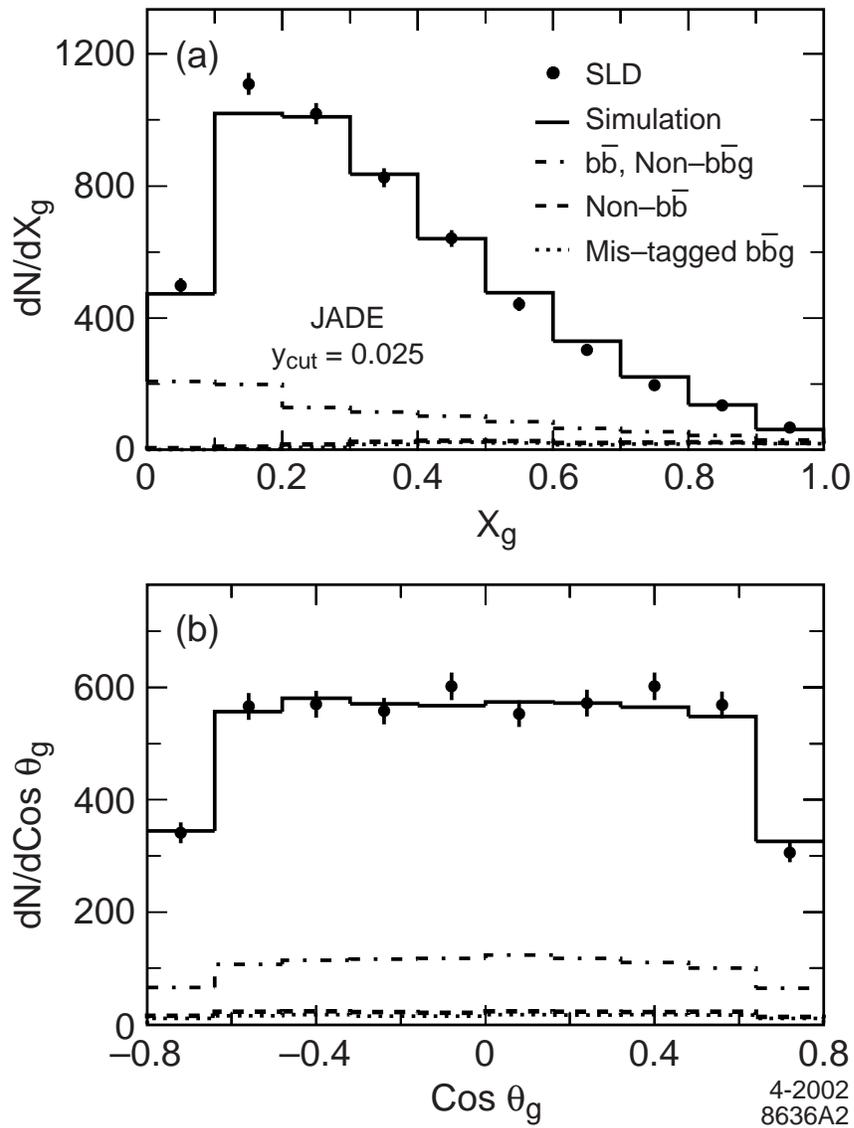}
\end{center}
\caption{Raw measured distributions of (a) $x_g$ and (b) cos$\theta_g$ (dots)
defined using the JADE algorithm; 
errors are statistical.
Histograms:  simulated distributions including background contributions.
}
\label{graw}
\end{figure}

For each algorithm the tagging purities, defined as the fraction of 
selected 3-jet events in which the two vertices were found in the
two jets containing the true $B$ hadrons,
are listed by gluon-jet number in 
Table~\ref{jetpur}.
We formed the distributions of two gluon-jet observables, 
the scaled energy $x_g = 2E_{\rm{gluon}}/\sqrt{s}$, and the polar angle 
w.r.t. the beamline, $\theta_g$.
For illustration, for the JADE algorithm the distributions are shown in 
Fig.~\ref{graw}; the simulation is also shown; it reproduces the data. 
Results for the other algorithms (not shown) are qualitatively similar.

The backgrounds were estimated using the simulation and are of three types:
non-$\bb$ events; $\bb$ but non-$\bbg$ events; and mis-tagged true $\bbg$
events.
Their contributions are shown in Fig.~\ref{graw} for the JADE case.
Results for the other algorithms (not shown) are qualitatively similar.
For each algorithm, the non-$\bb$ events make up roughly 1$\%$ of 
the selected sample and are dominated by
$c\overline{c}g$ events; roughly 70\% of these had the gluon jet correctly 
tagged, and the remainder comprises events in which the gluon split into a 
\cc or \bb, which yielded a real secondary vertex in the `wrong' jet.
Mis-tagged events, in which the gluon jet was mis-tagged as a $b$ or
$\bar{b}$-jet
and one of the true $b$- or $\bar{b}$-jets enters into the measured gluon 
distributions, comprise roughly 3\% of the sample; roughly two
thirds of these events contain a gluon splitting into \cc or \bb.
These two backgrounds are negligible except in the highest $x_g$ bin.

For all algorithms the dominant background (\eg for JADE,
roughly 16\% of the sample)
is formed by \bb but non-$\bbg$ events. These are true 
$\bb$ events that were not classified as 3-jet events at the parton level,
but were reconstructed and tagged as 3-jet $\bbg$ events in the detector
using the same jet algorithm and $y_{cut}$ value.
In a parton-level 2-jet event this can arise from the broadening of 
the particle flow
around the original $b$ and $\overline{b}$ directions due to hadronization
and weak decay; in particular, the relatively high-transverse-momentum 
$B$-decay products 
can cause the jet-finder to reconstruct a `fake' third jet,
which is almost always assigned as a (low-energy) gluon jet.
In addition, 
an event classified as 4-jet at the parton level may,
due to the overlap of their hadronization products, have
two of its jets combined in the detector by the jet-finder. 
In this case the combined jet is usually tagged as a gluon jet .
Since the calculations with which we compare below are not reliable
for 4-jet events, we consider such events to be a background.

\section{Correction of the Data}

For each algorithm, the distributions were corrected to obtain the true gluon
distributions $D^{true}(X)$ by applying a bin-by-bin procedure:
$D^{true}(X) = C(X)\;(D^{raw}(X)-B(X))$,
where $X$ = $x_g$ or cos$\theta_g$, $D^{raw}(X)$ 
is the raw distribution,
$B(X)$ is the background contribution, and
$C(X) \equiv D^{true}_{MC}(X)/D^{recon}_{MC}(X)$
is a correction that accounts for the
efficiency for accepting true \bbg events into the tagged sample, as well as
for bin-to-bin migrations caused by hadronization, 
the resolution of the detector, and bias of the jet-tagging technique.
Here $D^{true}_{MC}(X)$ is the true distribution for MC-generated \bbg events,
and $D^{recon}_{MC}(X)$ is the resulting distribution after full
simulation of the detector and application of the same analysis procedure
as applied to the data.

\begin{figure}[hbtp]
\begin{center}
\leavevmode
\epsfysize=15 cm.
\epsfbox{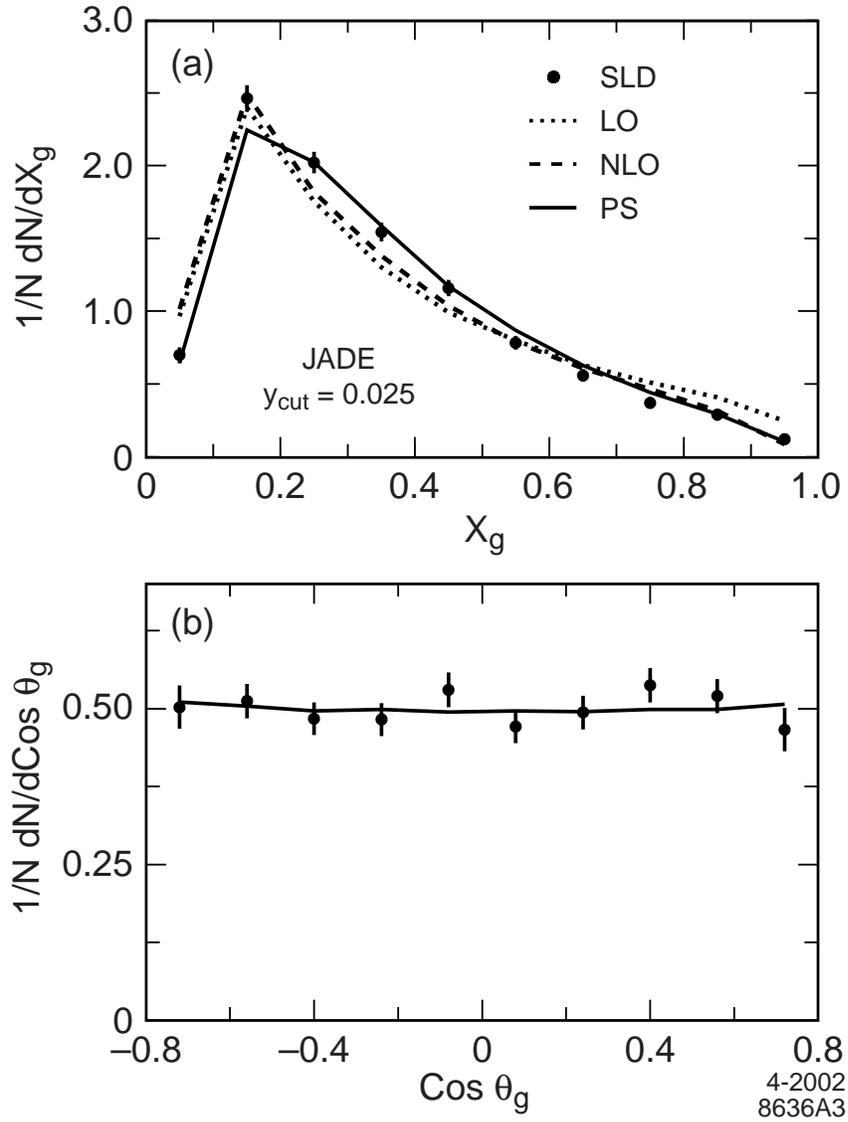}
\end{center}
\caption{Corrected distributions of (a) $x_g$ and (b) cos$\theta_g$ (dots)
defined using the JADE algorithm;
the error bars represent the sum in quadrature of the statistical and
systematic errors.
Perturbative QCD predictions (see text) are shown as lines joining entries
plotted at the respective bin centers.}
\label{gcorj}
\end{figure}

\begin{figure}[hbtp]
\begin{center}
\leavevmode
\epsfysize=15 cm.
\epsfbox{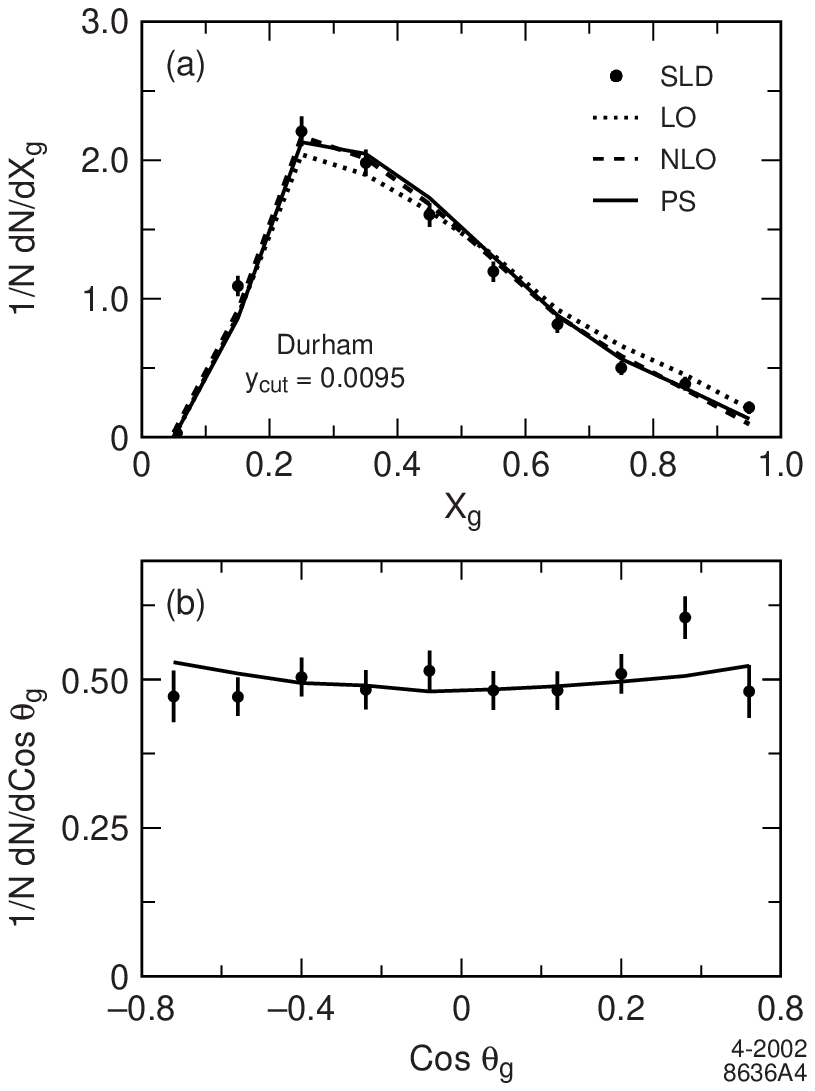}
\end{center}
\caption{Corrected distributions of (a) $x_g$ and (b) cos$\theta_g$ (dots);
defined using the Durham algorithm;
the error bars represent the sum in quadrature of the statistical and
systematic errors.
Perturbative QCD predictions (see text) are shown as lines joining entries
plotted at the respective bin centers.}
\label{gcord}
\end{figure}

\begin{figure}[hbtp]
\begin{center}
\leavevmode
\epsfysize=15 cm.
\epsfbox{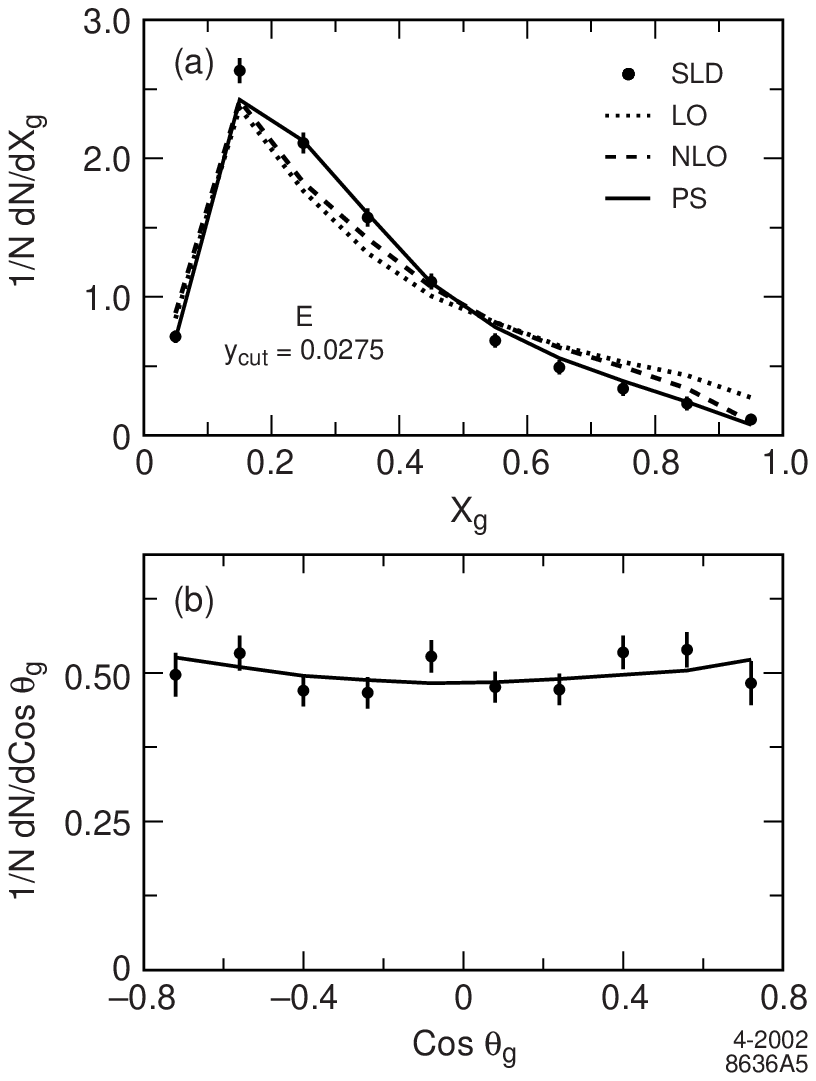}
\end{center}
\caption{Corrected distributions of (a) $x_g$ and (b) cos$\theta_g$ (dots);
defined using the E algorithm;
the error bars represent the sum in quadrature of the statistical and
systematic errors.
Perturbative QCD predictions (see text) are shown as lines joining entries
plotted at the respective bin centers.}
\label{gcore}
\end{figure}

\begin{figure}[hbtp]
\begin{center}
\leavevmode
\epsfysize=15 cm.
\epsfbox{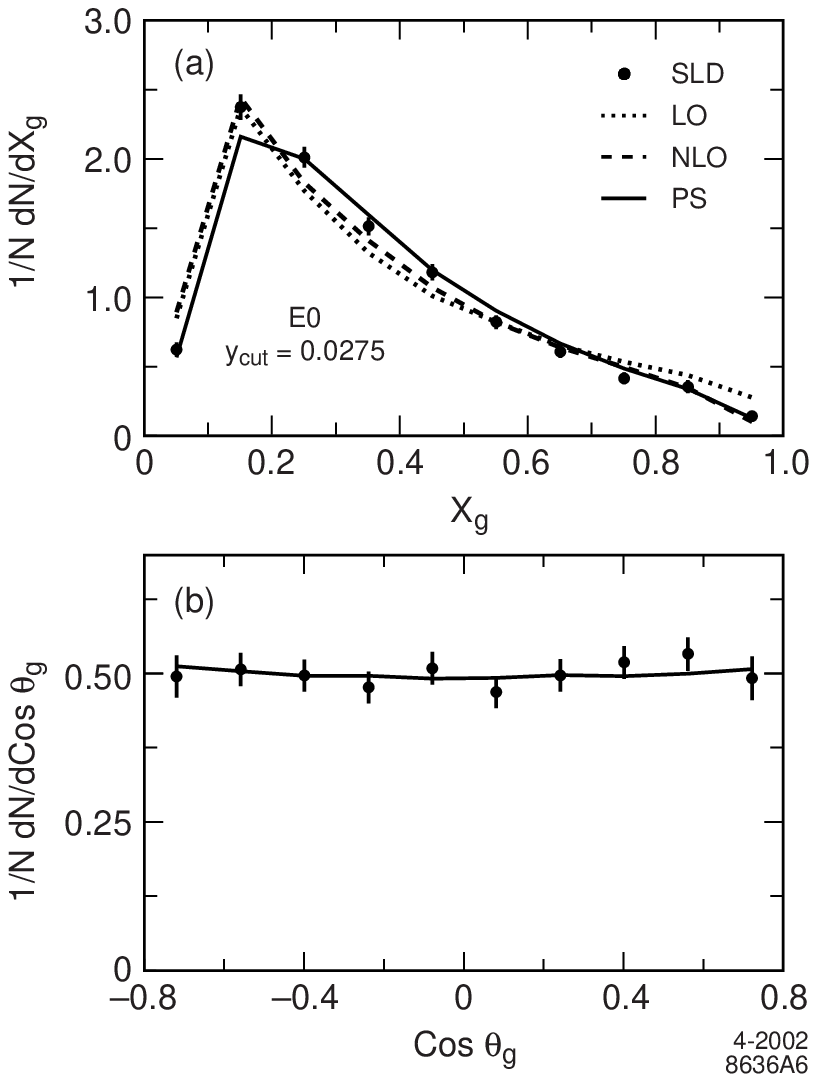}
\end{center}
\caption{Corrected distributions of (a) $x_g$ and (b) cos$\theta_g$ (dots);
defined using the E0 algorithm;
the error bars represent the sum in quadrature of the statistical and
systematic errors.
erturbative QCD predictions (see text) are shown as lines joining entries
plotted at the respective bin centers.}
\label{gcorez}
\end{figure}

\begin{figure}[hbtp]
\begin{center}
\leavevmode
\epsfysize=15 cm.
\epsfbox{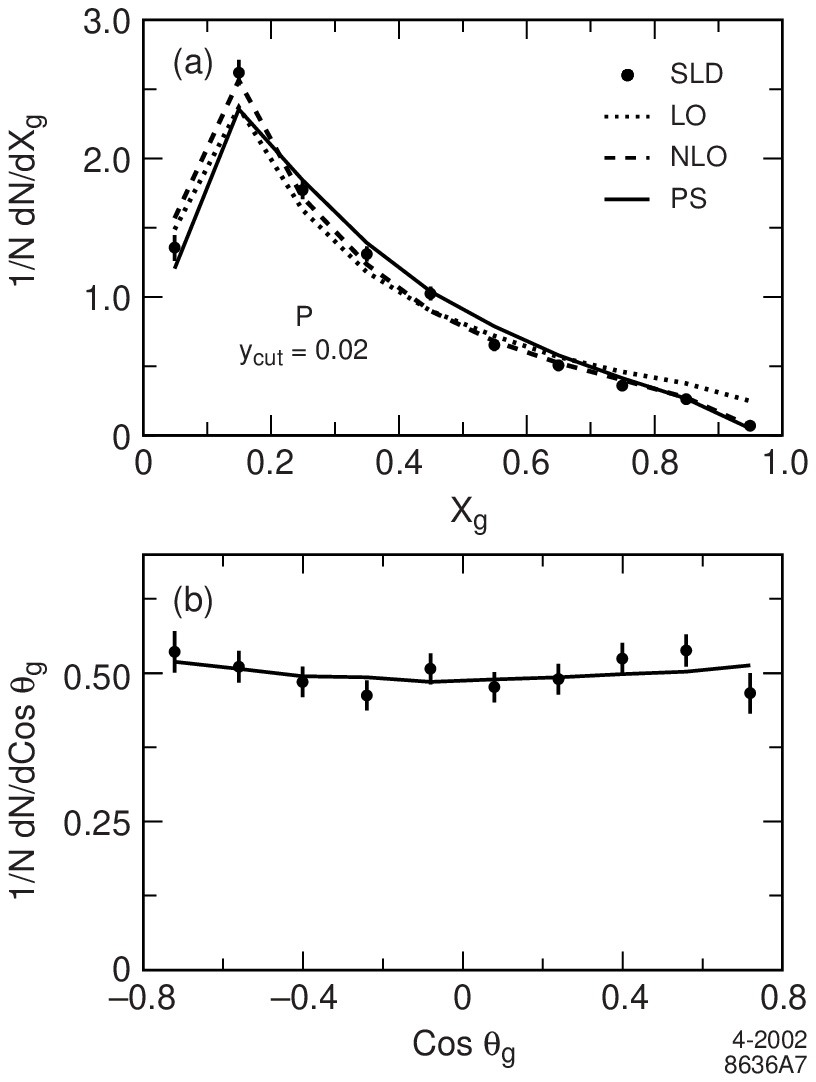}
\end{center}
\caption{Corrected distributions of (a) $x_g$ and (b) cos$\theta_g$ (dots);
defined using the P algorithm;
the error bars represent the sum in quadrature of the statistical and
systematic errors.
Perturbative QCD predictions (see text) are shown as lines joining entries
plotted at the respective bin centers.}
\label{gcorp}
\end{figure}

\begin{figure}[hbtp]
\begin{center}
\leavevmode
\epsfysize=15 cm.
\epsfbox{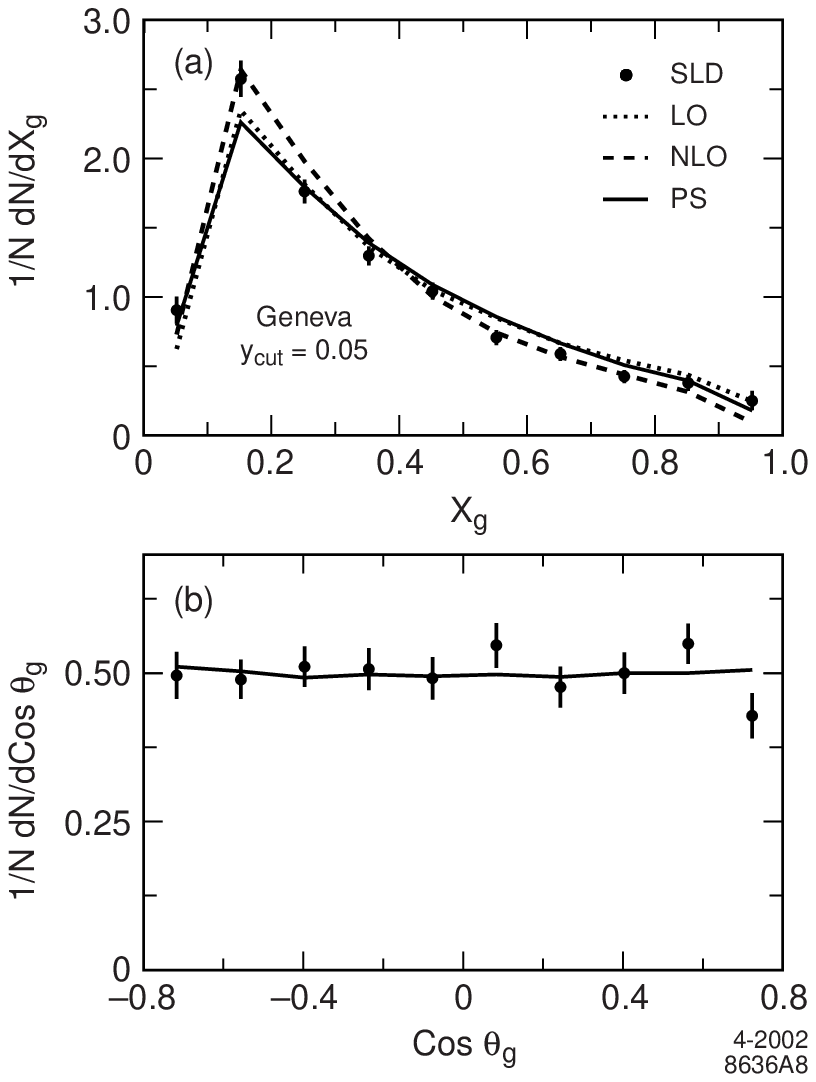}
\end{center}
\caption{Corrected distributions of (a) $x_g$ and (b) cos$\theta_g$ (dots);
defined using the Geneva algorithm;
the error bars represent the sum in quadrature of the statistical and
systematic errors.
Perturbative QCD predictions (see text) are shown as lines joining entries
plotted at the respective bin centers.}
\label{gcorg}
\end{figure}

The fully-corrected distributions are shown in
Figs.~\ref{gcorj}, \ref{gcord}, \ref{gcore}, \ref{gcorez}, \ref{gcorp}, and 
\ref{gcorg}. Since, in an
earlier study~\cite{sldflav}, we verified that the overall rate of \bbg-event production is
consistent with QCD expectations, we normalised the gluon distributions to
unit area and we study further the distribution shapes. 
In each case the peak in $x_g$ is a kinematic artefact of the jet-finding algorithm, 
which ensures that gluon jets are reconstructed with a non-zero energy,
and it depends on the $y_{cut}$ value.
The cos$\theta_g$ distributions are very nearly flat, in contrast to the
$1+\cos^2\theta$ behavior for quark jets.

We have considered sources of systematic uncertainty that potentially affect
our results.
These may be divided into uncertainties in modelling the detector and
uncertainties in the underlying physics modelling. 
To estimate the first case we systematically varied the track and event
selection requirements, as well as the track-finding
efficiency~\cite{sldflav,dervan}, the momentum and dip angle resolution, and 
the
probability of finding a fake vertex in a jet.
In the second case parameters used in our simulation, relating mainly to the
production and decay of charm and bottom hadrons, were varied within their
measurement errors~\cite{bfrag}. 
For each variation the data were recorrected to derive new $x_g$ and
cos$\theta_g$ distributions, and the deviation w.r.t. the standard case was
assigned as a systematic uncertainty. 
Although many of these variations affect the overall tagging efficiency, most
had little effect on the energy or polar angle dependence, and no variation
affects the conclusions below.
The largest contributions to the error arose from the measurement 
uncertainties on the probability
for gluon splitting into \bb ~(which dominates around $x_g$ $\sim$ 0.5)
or \cc ~(which dominates for $x_g$ $>$ 0.7).

All uncertainties were  conservatively assumed to be uncorrelated and were 
added in quadrature in each bin of $x_g$ and cos$\theta_g$. In any bin the
systematic error is typically much smaller than the statistical
error.
The data points with their total error bars are shown
in Figures~\ref{gcorj}, \ref{gcord}, \ref{gcore}, \ref{gcorez}, \ref{gcorp},
\ref{gcorg}; the data are listed in Tables~\ref{finalxg} and 
\ref{cosgfinal}.
In addition, as cross-checks, for each algorithm the $y_{cut}$ value
and the $M_{p_t}$ cut were varied around their respective default values; 
in no case did our conclusions change.

\begin{table}
\begin{center}
\begin{tabular}{|r@{--}l|r@{$\pm$}l@{$\pm$}l|
    r@{$\pm$}l@{$\pm$}l|r@{$\pm$}l@{$\pm$}l|} \hline
 \multicolumn{11}{|c|}{1/N dN/d$x_g$} \\
 \hline \hline
 \multicolumn{2}{|c|}{$x_g$ range} &  \multicolumn{3}{c|}{  JADE  } &
 \multicolumn{3}{c|}{ Durham } &  \multicolumn{3}{c|}{    E   } \\ \hline
 0.0 & 0.1 & 0.697& 0.055& 0.001 & 0.000& 0.000& 0.000 & 0.712& 0.045& 0.001 \\
 0.1 & 0.2 & 2.461& 0.093& 0.002 & 1.093& 0.073& 0.005 & 2.634& 0.091& 0.005 \\
 0.2 & 0.3 & 2.021& 0.074& 0.003 & 2.209& 0.106& 0.011 & 2.111& 0.075& 0.010 \\
 0.3 & 0.4 & 1.544& 0.064& 0.006 & 1.982& 0.093& 0.018 & 1.574& 0.065& 0.015 \\
 0.4 & 0.5 & 1.158& 0.056& 0.007 & 1.606& 0.083& 0.021 & 1.111& 0.055& 0.019 \\
 0.5 & 0.6 & 0.783& 0.047& 0.008 & 1.196& 0.072& 0.015 & 0.684& 0.045& 0.024 \\
 0.6 & 0.7 & 0.559& 0.042& 0.008 & 0.814& 0.060& 0.012 & 0.493& 0.041& 0.029 \\
 0.7 & 0.8 & 0.371& 0.038& 0.009 & 0.501& 0.049& 0.010 & 0.337& 0.039& 0.033 \\
 0.8 & 0.9 & 0.287& 0.038& 0.012 & 0.385& 0.048& 0.011 & 0.231& 0.038& 0.032 \\
 0.9 & 1.0 & 0.119& 0.028& 0.013 & 0.214& 0.043& 0.013 & 0.114& 0.026& 0.020 \\
 \hline\hline
 \multicolumn{2}{|c|}{$x_g$ range} &  \multicolumn{3}{c|}{   E0   } &
 \multicolumn{3}{c|}{    P   } &  \multicolumn{3}{c|}{ Geneva } \\ \hline
 0.0 & 0.1 & 0.620& 0.050& 0.001 & 1.362& 0.092& 0.001 & 0.912& 0.096& 0.004 \\
 0.1 & 0.2 & 2.369& 0.092& 0.002 & 2.627& 0.092& 0.003 & 2.582& 0.130& 0.008 \\
 0.2 & 0.3 & 2.007& 0.076& 0.004 & 1.779& 0.064& 0.004 & 1.768& 0.085& 0.007 \\
 0.3 & 0.4 & 1.509& 0.065& 0.006 & 1.317& 0.054& 0.006 & 1.304& 0.070& 0.007 \\
 0.4 & 0.5 & 1.176& 0.058& 0.007 & 1.033& 0.049& 0.007 & 1.049& 0.061& 0.006 \\
 0.5 & 0.6 & 0.817& 0.050& 0.007 & 0.659& 0.041& 0.007 & 0.713& 0.052& 0.007 \\
 0.6 & 0.7 & 0.604& 0.045& 0.008 & 0.511& 0.039& 0.008 & 0.596& 0.049& 0.005 \\
 0.7 & 0.8 & 0.411& 0.041& 0.009 & 0.367& 0.036& 0.009 & 0.433& 0.047& 0.008 \\
 0.8 & 0.9 & 0.351& 0.044& 0.014 & 0.268& 0.036& 0.012 & 0.386& 0.055& 0.015 \\
 0.9 & 1.0 & 0.137& 0.033& 0.016 & 0.076& 0.016& 0.008 & 0.259& 0.064& 0.028 \\
 \hline
\end{tabular}
\end{center}
\caption{Fully-corrected differential cross-sections
for hard gluon-jet production as a function of jet
energy $x_g$.  The first error is statistical, the
second systematic}
\label{finalxg}
\end{table}


\begin{table}
\begin{center}
\begin{tabular}{|r@{--}l|r@{$\pm$}l@{$\pm$}l|
    r@{$\pm$}l@{$\pm$}l|r@{$\pm$}l@{$\pm$}l|} \hline
 \multicolumn{11}{|c|}{1/N dN/dcos$\theta_g$} \\
 \hline \hline
 \multicolumn{2}{|c|}{cos$\theta_g$ range} &  \multicolumn{3}{c|}{  JADE  } &
 \multicolumn{3}{c|}{ Durham } &  \multicolumn{3}{c|}{    E   } \\ \hline
 -0.80 & -0.64 & 0.503& 0.035& 0.004& 0.471& 0.043& 0.006& 0.497& 0.036& 0.010 \\
 -0.64 & -0.48 & 0.512& 0.027& 0.003& 0.471& 0.032& 0.005& 0.533& 0.028& 0.009 \\
 -0.48 & -0.32 & 0.484& 0.026& 0.003& 0.504& 0.032& 0.006& 0.470& 0.025& 0.009 \\
 -0.32 & -0.16 & 0.482& 0.027& 0.003& 0.483& 0.033& 0.006& 0.467& 0.025& 0.009 \\
 -0.16 & 0.0   & 0.530& 0.028& 0.003& 0.515& 0.033& 0.006& 0.528& 0.026& 0.009 \\
   0.0 & 0.16  & 0.471& 0.027& 0.003& 0.481& 0.032& 0.005& 0.476& 0.025& 0.008 \\
  0.16 & 0.32  & 0.494& 0.027& 0.003& 0.481& 0.032& 0.005& 0.472& 0.025& 0.008 \\
  0.32 & 0.48  & 0.537& 0.028& 0.003& 0.509& 0.033& 0.005& 0.535& 0.027& 0.008 \\
  0.48 & 0.64  & 0.520& 0.027& 0.003& 0.604& 0.035& 0.006& 0.539& 0.028& 0.009 \\
  0.64 & 0.80  & 0.466& 0.035& 0.003& 0.480& 0.044& 0.005& 0.483& 0.036& 0.009 \\
 \hline\hline
 \multicolumn{2}{|c|}{cos$\theta_g$ range} &  \multicolumn{3}{c|}{   E0   } &
 \multicolumn{3}{c|}{    P   } &  \multicolumn{3}{c|}{ Geneva } \\ \hline
 -0.80 & -0.64 & 0.496& 0.035& 0.003& 0.536& 0.035& 0.004& 0.496& 0.039& 0.003 \\
 -0.64 & -0.48 & 0.508& 0.028& 0.003& 0.511& 0.027& 0.003& 0.490& 0.033& 0.004 \\
 -0.48 & -0.32 & 0.497& 0.027& 0.003& 0.486& 0.025& 0.004& 0.511& 0.034& 0.004 \\
 -0.32 & -0.16 & 0.477& 0.027& 0.003& 0.463& 0.025& 0.003& 0.507& 0.035& 0.004 \\
 -0.16 & 0.0   & 0.510& 0.028& 0.003& 0.508& 0.026& 0.003& 0.492& 0.036& 0.003 \\
   0.0 & 0.16  & 0.469& 0.027& 0.003& 0.477& 0.025& 0.003& 0.547& 0.037& 0.004 \\
  0.16 & 0.32  & 0.497& 0.027& 0.003& 0.490& 0.026& 0.003& 0.477& 0.034& 0.004 \\
  0.32 & 0.48  & 0.519& 0.028& 0.003& 0.525& 0.027& 0.003& 0.501& 0.035& 0.004 \\
  0.48 & 0.64  & 0.533& 0.028& 0.003& 0.538& 0.027& 0.003& 0.550& 0.034& 0.004 \\
  0.64 & 0.80  & 0.493& 0.037& 0.004& 0.466& 0.034& 0.004& 0.429& 0.038& 0.003 \\
 \hline
\end{tabular}
\end{center}
\caption{Fully corrected differential cross-sections
for hard gluon-jet production as a function of jet
polar angle, cos$\theta_g$.  The first error is statistical, the
second systematic}
\label{cosgfinal}
\end{table}

\section{Comparison with QCD Predictions}

We compared the data with perturbative QCD predictions for the respective jet
algorithm and $y_{cut}$ value.
We calculated leading-order (LO) and next-to-LO (NLO) predictions using JETSET.
We also derived these distributions using the `parton  shower' (PS) implemented
in JETSET;
this is operationally equivalent to a calculation in which leading 
and next-to-leading ln$y_c$ terms are partially resummed to all orders in \alp.
In physical terms this allows events to be generated with multiple orders of
parton radiation, in contrast to the maximum number of 3 (4) partons allowed in
the LO~(NLO) calculations, respectively.
Configurations with $\geq3$ partons are relevant to the observables considered
here since they may be resolved as 3-jet events by the jet-finding algorithm.
These predictions are shown in 
Figs.~\ref{gcorj}, ~\ref{gcord}, ~\ref{gcore}, ~\ref{gcorez}, ~\ref{gcorp} and 
\ref{gcorg}.

In the case of the cos$\theta_g$ distributions 
the three calculations are indistinguishable 
and they reproduce the data. For clarity we show in
Figs.~\ref{gcorj}, \ref{gcord}, \ref{gcore}, \ref{gcorez}, \ref{gcorp}, and
\ref{gcorg} only the PS calculations.
We conclude that the cos$\theta_g$ observable is insensitive to 
the details of higher order soft parton emission.

In the case of $x_g$, for the JADE, E, E0 and P algorithms 
the LO calculation reproduces the 
main features of the shape of the
distribution, but it yields too few events in the region $0.2<x_g<0.5$, and too
many events for $x_g<0.1$ and $x_g>0.6$.
The NLO calculation shows qualitatively similar behavior, although it 
reproduces the data noticeably better, especially for $x_g>0.6$.
In the case of the JADE, E and E0 algorithms
the PS calculation provides the best description of the data across the full 
$x_g$ range, although it tends to underestimate the height of the kinematic
peak; in the case of the P algorithm the PS calculation is slightly
worse than the NLO calculation.
These results suggest that the data are sensitive to 
multiple orders of parton radiation, the details of which need to be
included in the perturbative QCD calculation.
This is in agreement with our earlier inclusive measurement of jet energy 
distributions (for the JADE algorithm only) using flavor-inclusive 
\z0 decays~\cite{hwang}. 

In the case of $x_g$ defined using the Geneva algorithm (Fig.~\ref{gcorg}), 
there are clear differences among the three calculations, but
the NLO calculation reproduces the data best. 
Finally, in the case of the Durham algorithm (Fig.~\ref{gcord}),
the differences among the three calculations are relatively small,
and both the NLO and PS calculations provide a good description of
the data. This is consistent with the original motivation for the Durham 
algorithm~\cite{durham}, which was explicitly designed 
to yield a jet structure that is relatively insensitive to the presence of
additional soft partons.

We conclude that perturbative QCD in the PS (JADE, Durham, E, E0 algorithms)
or NLO (P, Durham, Geneva algorithms) approximation reproduces the gluon
distributions in \bbg events.
However, it is interesting to consider the extent to which anomalous
chromomagnetic contributions are allowed by the data.
The Lagrangian represented by Eq.~\ref{lag} yields a model that is
non-renormalizable.
Nevertheless tree-level predictions can be derived \cite{tom1} and used for a
`straw man' comparison with QCD. 
For each jet algorithm, in each bin of the $x_g$ distribution, 
we parametrised the leading-order effect
of an anomalous chromomagnetic moment and added it to the PS calculation to
arrive at an effective QCD prediction including the anomalous moment at
leading-order.
A $\chi^2$ minimization fit was performed to the data with $\kappa$ as a free
parameter. The corresponding $\kappa$ and $\chi^2$ values are shown in 
Table~\ref{chisq}.
In all cases $\kappa$ is consistent with zero.
For each algorithm the confidence level of the fit is smaller
than the confidence level based on the $\chi^2$ for the
comparison with the standard PS calculation.
We conclude that our data show no evidence for any beyond-QCD effects,
and we set 95\% confidence-level (C.L.)
limits on $\kappa$; these are shown in Table~\ref{chisq}.
Since the results are highly correlated, we quote best limits on $\kappa$ using
the JADE algorithm, yielding 
$-0.058 < \kappa < 0.043$ at the 95$\%$ C.L.     
            
\begin{table} [t]
\begin{center}
\begin{tabular}{|l|c|c|c|} \hline
Jet algorithm & $\kappa$ & $\chi^2$ (10 bins) & 95\% C.L. limits \\
\hline
JADE    & $-0.008\pm0.026$   &  15.9  & $-0.058<\kappa<0.043$  \\
Durham  & $0.020\pm0.043$    &  21.8  & $-0.065<\kappa<0.106$  \\
E       & $-0.005\pm0.028$   &  13.6  & $-0.060<\kappa<0.050$  \\
E0      & $-0.006\pm0.027$   &  15.6  & $-0.060<\kappa<0.047$  \\
P       & $-0.002\pm0.025$   &  31.7  & $-0.052<\kappa<0.047$  \\
Geneva  & $-0.006\pm0.026$   &  24.4  & $-0.056<\kappa<0.045$  \\
\hline
\end{tabular}
\end{center}
\caption{Best-fit $\kappa$ values and 95\% C.L. limits.
}
\label{chisq}
\end{table}
 
\section{Conclusion}

In conclusion, we used the precise SLD tracking system to tag the gluon in
3-jet $e^+e^-\rightarrow Z^0\rightarrow\bbg$ events. We studied the structure
of these events in terms of the scaled gluon energy and polar angle, measured 
across the full kinematic range.
We compared our data with perturbative QCD predictions and found that
beyond-LO QCD contributions are needed to describe the energy distribution.
We also investigated an anomalous $b$-quark chromomagnetic moment, $\kappa$,
which would affect the shape of the energy distribution.
We set 95$\%$ c.l. limits of $-0.06 < \kappa < 0.04$.
These results are consistent with, more precise than, and supersede
those in our earlier publication~\cite{sldbbg}.

\vskip .5truecm

We thank the personnel of the SLAC accelerator department and the technical
staffs of our collaborating institutions for their outstanding efforts
on our behalf.
We thank A.~Brandenburg, P.~Uwer and T.~Rizzo for many helpful discussions.

\vskip 1truecm

\vbox{
\footnotesize\renewcommand{\baselinestretch}{1}
\noindent
 This work was supported by Department of Energy
  contracts:
  DE-FG02-91ER40676 (BU),
  DE-FG03-91ER40618 (UCSB),
  DE-FG03-92ER40689 (UCSC),
  DE-FG03-93ER40788 (CSU),
  DE-FG02-91ER40672 (Colorado),
  DE-FG02-91ER40677 (Illinois),
  DE-AC03-76SF00098 (LBL),
  DE-FG02-92ER40715 (Massachusetts),
  DE-FC02-94ER40818 (MIT),
  DE-FG03-96ER40969 (Oregon),
  DE-AC03-76SF00515 (SLAC),
  DE-FG05-91ER40627 (Tennessee),
  DE-FG02-95ER40896 (Wisconsin),
  DE-FG02-92ER40704 (Yale);
  National Science Foundation grants:
  PHY-91-13428 (UCSC),
  PHY-89-21320 (Columbia),
  PHY-92-04239 (Cincinnati),
  PHY-95-10439 (Rutgers),
  PHY-88-19316 (Vanderbilt),
  PHY-92-03212 (Washington);
  the UK Particle Physics and Astronomy Research Council
  (Brunel, Oxford and RAL);
  the Istituto Nazionale di Fisica Nucleare of Italy
  (Bologna, Ferrara, Frascati, Pisa, Padova, Perugia);
  the Japan-US Cooperative Research Project on High Energy Physics
  (Nagoya, Tohoku);
  and the Korea Science and Engineering Foundation (Soongsil).}

\vfill
\eject

\section*{$^{**}$List of Authors}
%
%
%
\begin{center}
\def\iAOMORI{$^{(1)}$}
\def\iBRI{$^{(2)}$}
\def\iBRUN{$^{(3)}$}
\def\iBU{$^{(4)}$}
\def\iCOLO{$^{(5)}$}
\def\iCSU{$^{(6)}$}
\def\iFERR{$^{(7)}$}
\def\iFRAS{$^{(8)}$}
\def\iJHU{$^{(9)}$}
\def\iLBL{$^{(10)}$}
\def\iMASS{$^{(11)}$}
\def\iMISSI{$^{(12)}$}
\def\iMIT{$^{(13)}$}
\def\iMOSCOW{$^{(14)}$}
\def\iNAGO{$^{(15)}$}
\def\iOREG{$^{(16)}$}
\def\iOXF{$^{(17)}$}
\def\iPERU{$^{(18)}$}
\def\iRAL{$^{(19)}$}
\def\iRUTG{$^{(20)}$}
\def\iSLAC{$^{(21)}$}
\def\iSOONG{$^{(22)}$}
\def\iTENN{$^{(23)}$}
\def\iTOHO{$^{(24)}$}
\def\iUCSB{$^{(25)}$}
\def\iUCSC{$^{(26)}$}
\def\iVAND{$^{(27)}$}
\def\iWASH{$^{(28)}$}
\def\iWISC{$^{(29)}$}
\def\iYALE{$^{(30)}$}

  \baselineskip=.75\baselineskip
\mbox{Koya Abe\unskip,\iTOHO}
\mbox{Kenji Abe\unskip,\iNAGO}
\mbox{T. Abe\unskip,\iSLAC}
\mbox{I. Adam\unskip,\iSLAC}
\mbox{H. Akimoto\unskip,\iSLAC}
\mbox{D. Aston\unskip,\iSLAC}
\mbox{K.G. Baird\unskip,\iMASS}
\mbox{C. Baltay\unskip,\iYALE}
\mbox{H.R. Band\unskip,\iWISC}
\mbox{T.L. Barklow\unskip,\iSLAC}
\mbox{J.M. Bauer\unskip,\iMISSI}
\mbox{G. Bellodi\unskip,\iOXF}
\mbox{R. Berger\unskip,\iSLAC}
\mbox{G. Blaylock\unskip,\iMASS}
\mbox{J.R. Bogart\unskip,\iSLAC}
\mbox{G.R. Bower\unskip,\iSLAC}
\mbox{J.E. Brau\unskip,\iOREG}
\mbox{M. Breidenbach\unskip,\iSLAC}
\mbox{W.M. Bugg\unskip,\iTENN}
\mbox{D. Burke\unskip,\iSLAC}
\mbox{T.H. Burnett\unskip,\iWASH}
\mbox{P.N. Burrows\unskip,\iOXF}
\mbox{A. Calcaterra\unskip,\iFRAS}
\mbox{R. Cassell\unskip,\iSLAC}
\mbox{A. Chou\unskip,\iSLAC}
\mbox{H.O. Cohn\unskip,\iTENN}
\mbox{J.A. Coller\unskip,\iBU}
\mbox{M.R. Convery\unskip,\iSLAC}
\mbox{V. Cook\unskip,\iWASH}
\mbox{R.F. Cowan\unskip,\iMIT}
\mbox{G. Crawford\unskip,\iSLAC}
\mbox{C.J.S. Damerell\unskip,\iRAL}
\mbox{M. Daoudi\unskip,\iSLAC}
\mbox{N. de Groot\unskip,\iBRI}
\mbox{R. de Sangro\unskip,\iFRAS}
\mbox{D.N. Dong\unskip,\iMIT}
\mbox{M. Doser\unskip,\iSLAC}
\mbox{R. Dubois\unskip,}
\mbox{I. Erofeeva\unskip,\iMOSCOW}
\mbox{V. Eschenburg\unskip,\iMISSI}
\mbox{E. Etzion\unskip,\iWISC}
\mbox{S. Fahey\unskip,\iCOLO}
\mbox{D. Falciai\unskip,\iFRAS}
\mbox{J.P. Fernandez\unskip,\iUCSC}
\mbox{K. Flood\unskip,\iMASS}
\mbox{R. Frey\unskip,\iOREG}
\mbox{E.L. Hart\unskip,\iTENN}
\mbox{K. Hasuko\unskip,\iTOHO}
\mbox{S.S. Hertzbach\unskip,\iMASS}
\mbox{M.E. Huffer\unskip,\iSLAC}
\mbox{X. Huynh\unskip,\iSLAC}
\mbox{M. Iwasaki\unskip,\iOREG}
\mbox{D.J. Jackson\unskip,\iRAL}
\mbox{P. Jacques\unskip,\iRUTG}
\mbox{J.A. Jaros\unskip,\iSLAC}
\mbox{Z.Y. Jiang\unskip,\iSLAC}
\mbox{A.S. Johnson\unskip,\iSLAC}
\mbox{J.R. Johnson\unskip,\iWISC}
\mbox{R. Kajikawa\unskip,\iNAGO}
\mbox{M. Kalelkar\unskip,\iRUTG}
\mbox{H.J. Kang\unskip,\iRUTG}
\mbox{R.R. Kofler\unskip,\iMASS}
\mbox{R.S. Kroeger\unskip,\iMISSI}
\mbox{M. Langston\unskip,\iOREG}
\mbox{D.W.G. Leith\unskip,\iSLAC}
\mbox{V. Lia\unskip,\iMIT}
\mbox{C. Lin\unskip,\iMASS}
\mbox{G. Mancinelli\unskip,\iRUTG}
\mbox{S. Manly\unskip,\iYALE}
\mbox{G. Mantovani\unskip,\iPERU}
\mbox{T.W. Markiewicz\unskip,\iSLAC}
\mbox{T. Maruyama\unskip,\iSLAC}
\mbox{A.K. McKemey\unskip,\iBRUN}
\mbox{R. Messner\unskip,\iSLAC}
\mbox{K.C. Moffeit\unskip,\iSLAC}
\mbox{T.B. Moore\unskip,\iYALE}
\mbox{M. Morii\unskip,\iSLAC}
\mbox{D. Muller\unskip,\iSLAC}
\mbox{V. Murzin\unskip,\iMOSCOW}
\mbox{S. Narita\unskip,\iTOHO}
\mbox{U. Nauenberg\unskip,\iCOLO}
\mbox{H. Neal\unskip,\iYALE}
\mbox{G. Nesom\unskip,\iOXF}
\mbox{N. Oishi\unskip,\iNAGO}
\mbox{D. Onoprienko\unskip,\iTENN}
\mbox{L.S. Osborne\unskip,\iMIT}
\mbox{R.S. Panvini\unskip,\iVAND}
\mbox{C.H. Park\unskip,\iSOONG}
\mbox{I. Peruzzi\unskip,\iFRAS}
\mbox{M. Piccolo\unskip,\iFRAS}
\mbox{L. Piemontese\unskip,\iFERR}
\mbox{R.J. Plano\unskip,\iRUTG}
\mbox{R. Prepost\unskip,\iWISC}
\mbox{C.Y. Prescott\unskip,\iSLAC}
\mbox{B.N. Ratcliff\unskip,\iSLAC}
\mbox{J. Reidy\unskip,\iMISSI}
\mbox{P.L. Reinertsen\unskip,\iUCSC}
\mbox{L.S. Rochester\unskip,\iSLAC}
\mbox{P.C. Rowson\unskip,\iSLAC}
\mbox{J.J. Russell\unskip,\iSLAC}
\mbox{O.H. Saxton\unskip,\iSLAC}
\mbox{T. Schalk\unskip,\iUCSC}
\mbox{B.A. Schumm\unskip,\iUCSC}
\mbox{J. Schwiening\unskip,\iSLAC}
\mbox{V.V. Serbo\unskip,\iSLAC}
\mbox{G. Shapiro\unskip,\iLBL}
\mbox{N.B. Sinev\unskip,\iOREG}
\mbox{J.A. Snyder\unskip,\iYALE}
\mbox{H. Staengle\unskip,\iCSU}
\mbox{A. Stahl\unskip,\iSLAC}
\mbox{P. Stamer\unskip,\iRUTG}
\mbox{H. Steiner\unskip,\iLBL}
\mbox{D. Su\unskip,\iSLAC}
\mbox{F. Suekane\unskip,\iTOHO}
\mbox{A. Sugiyama\unskip,\iNAGO}
\mbox{A. Suzuki\unskip,\iNAGO}
\mbox{M. Swartz\unskip,\iJHU}
\mbox{F.E. Taylor\unskip,\iMIT}
\mbox{J. Thom\unskip,\iSLAC}
\mbox{E. Torrence\unskip,\iMIT}
\mbox{T. Usher\unskip,\iSLAC}
\mbox{J. Va'vra\unskip,\iSLAC}
\mbox{R. Verdier\unskip,\iMIT}
\mbox{D.L. Wagner\unskip,\iCOLO}
\mbox{A.P. Waite\unskip,\iSLAC}
\mbox{S. Walston\unskip,\iOREG}
\mbox{A.W. Weidemann\unskip,\iTENN}
\mbox{E.R. Weiss\unskip,\iWASH}
\mbox{J.S. Whitaker\unskip,\iBU}
\mbox{S.H. Williams\unskip,\iSLAC}
\mbox{S. Willocq\unskip,\iMASS}
\mbox{R.J. Wilson\unskip,\iCSU}
\mbox{W.J. Wisniewski\unskip,\iSLAC}
\mbox{J.L. Wittlin\unskip,\iMASS}
\mbox{M. Woods\unskip,\iSLAC}
\mbox{T.R. Wright\unskip,\iWISC}
\mbox{R.K. Yamamoto\unskip,\iMIT}
\mbox{J. Yashima\unskip,\iTOHO}
\mbox{S.J. Yellin\unskip,\iUCSB}
\mbox{C.C. Young\unskip,\iSLAC}
\mbox{H. Yuta\unskip.\iAOMORI}

\it
  \vskip \baselineskip                   
  \baselineskip=.75\baselineskip   
\iAOMORI
  Aomori University, Aomori, 030 Japan, \break
\iBRI
  University of Bristol, Bristol, United Kingdom, \break
\iBRUN
  Brunel University, Uxbridge, Middlesex, UB8 3PH United Kingdom, \break
\iBU
  Boston University, Boston, Massachusetts 02215, \break
\iCOLO
  University of Colorado, Boulder, Colorado 80309, \break
\iCSU
  Colorado State University, Ft. Collins, Colorado 80523, \break
\iFERR
  INFN Sezione di Ferrara and Universita di Ferrara, I-44100 Ferrara, Italy,
\break
\iFRAS
  INFN Laboratori Nazionali di Frascati, I-00044 Frascati, Italy, \break
\iJHU
  Johns Hopkins University,  Baltimore, Maryland 21218-2686, \break
\iLBL
  Lawrence Berkeley Laboratory, University of California, Berkeley, California
94720, \break
\iMASS
  University of Massachusetts, Amherst, Massachusetts 01003, \break
\iMISSI
  University of Mississippi, University, Mississippi 38677, \break
\iMIT
  Massachusetts Institute of Technology, Cambridge, Massachusetts 02139, \break
\iMOSCOW
  Institute of Nuclear Physics, Moscow State University, 119899 Moscow, Russia,
\break
\iNAGO
  Nagoya University, Chikusa-ku, Nagoya, 464 Japan, \break
\iOREG
  University of Oregon, Eugene, Oregon 97403, \break
\iOXF
  Oxford University, Oxford, OX1 3RH, United Kingdom, \break
\iPERU
  INFN Sezione di Perugia and Universita di Perugia, I-06100 Perugia, Italy,
\break
\iRAL
  Rutherford Appleton Laboratory, Chilton, Didcot, Oxon OX11 0QX United Kingdom,
\break
\iRUTG
  Rutgers University, Piscataway, New Jersey 08855, \break
\iSLAC
  Stanford Linear Accelerator Center, Stanford University, Stanford, California
94309, \break
\iSOONG
  Soongsil University, Seoul, Korea 156-743, \break
\iTENN
  University of Tennessee, Knoxville, Tennessee 37996, \break
\iTOHO
  Tohoku University, Sendai, 980 Japan, \break
\iUCSB
  University of California at Santa Barbara, Santa Barbara, California 93106,
\break
\iUCSC
  University of California at Santa Cruz, Santa Cruz, California 95064, \break
\iVAND
  Vanderbilt University, Nashville,Tennessee 37235, \break
\iWASH
  University of Washington, Seattle, Washington 98105, \break
\iWISC
  University of Wisconsin, Madison,Wisconsin 53706, \break
\iYALE
  Yale University, New Haven, Connecticut 06511. \break

\rm
%

\end{center}


\begin{thebibliography}{99}
\bibitem{quark} 
We do not distinguish between particle and antiparticle.
\end{thebibliography}
\end{document}